\documentclass[11pt,aps,preprint,epsfig,showpacs]{revtex4}
\usepackage{subfigure}
\usepackage{graphicx}
\usepackage{epsfig}
\begin{document}
\title{Statistical Mechanics of  DNA Rupture: Theory and Simulations}
\author{S. Nath, T. Modi$^{1}$, R. K. Mishra, D. Giri$^{1}$, B. P. Mandal and S. Kumar} 
\affiliation{Department of Physics, Banaras Hindu University,
     Varanasi 221 005, India \\
$^{1}$~Department of Physics, Indian Institute of Technology (BHU), Varanasi 221005, India}
\begin{abstract}
We study the effects of the shear force on the rupture mechanism  on  
a double stranded DNA.  Motivated by recent experiments, 
we perform the atomistic simulations with explicit solvent to obtain
the distributions of extension in hydrogen and covalent bonds below
the rupture force. We obtain a significant difference between the atomistic 
simulations and the existing results in the literature based on the 
coarse-grained models (theory and simulations).  We discuss the 
possible reasons and improve the coarse-grained model by incorporating
the consequences of semi-microscopic details of the nucleotides in its description.
The distributions obtained by the modified model (simulations and theoretical) are 
qualitatively similar to the  one obtained using  atomistic simulations.

\end{abstract}
\maketitle

\section{introduction}
Single molecule force spectroscopy (SMFS) techniques have  
enhanced our understanding about the inter- and intra- molecular 
interactions involved in the stability of DNA and biological processes 
{\it e.g.} transcription, replication, slippage, rupture etc. 
\cite{albert,israel,kumarphys,Bockelmann, Bock,Lee_Science94,Strunge,Irina,prentiss1,hatch,Cludia,gaub,nik13}. 
Initially, it was thought that the interactions detected in SMFS experiments
would be mostly of a mechanical nature and can be calculated by
knowing the value of the applied force. However, insights gathered from these
experiments revealed that the measurement of molecular interactions depends not only 
on the magnitude of the applied force, but also on how and where 
the force is applied \cite{kumarphys,Bockelmann,Bock,Lee_Science94,Strunge,Irina,prentiss1,
hatch,Cludia,gaub,nik13}. For example, Bockelmamm and coworkers \cite{Bockelmann, Bock} 
applied the  force perpendicular to the helix direction (DNA unzipping) and measured the 
unzipping force $\sim 15$ pN, whereas, Lee et al \cite{Lee_Science94} studied 
the unbinding of double stranded DNA (dsDNA) by applying a force
along the helix direction (rupture of DNA), and measured the rupture force, which is one order magnitude
greater than the unzipping force. Strunz {\it et al} \cite{Strunge,Irina} investigated the unbinding 
of DNA duplex of various lengths and found that the unbinding force depends 
on the loading rate and sequence length. It was also found that 
changing the pulling direction results different unbinding forces \cite{Cludia,nik13}.

By  expressing the bond energy and base-pairing energy in the form of harmonic 
oscillators in the ladder model of dsDNA (homosequence) of length $N$ base-pairs, 
de Gennes \cite{degennes} proposed the maximum force required for the rupture is
\begin{equation}
F_c= 2 f_1 (\chi^{-1} \tanh(\chi \frac{N}{2})),
\end{equation}
where $f_1 $ is the force required to separate a single base-pair and 
$\chi^{-1} = \sqrt{Q/2R}$ is the de Gennes characteristic length. Here, $Q$ 
and $R$ are the spring constants of covalent (backbone) and hydrogen bonds, respectively. 
Eq. 1 predicts that the rupture force increases
linearly with length for small values of $N$ and saturates at the higher values
of $N$, which is consistent with recent experiment \cite{hatch}.  In an another study, 
Chakrabarti and Nelson \cite{nelson} extended the de Gennes model (nonlinear 
generalization of the ladder model) and studied the effects of sequence heterogeneity. 
Mishra {\it et al} \cite{Mishra} considered a homosequence of DNA, where the covalent bonds
and base-pairing interactions are modeled by the harmonic spring and  
Lennard-Jones (LJ) potential, respectively. Using Langevin dynamics (LD) simulations 
\cite{Allen, Smith}, they obtained the distribution of stretching of hydrogen bonds 
and the extension in the covalent bonds for a wide range of forces below the rupture, 
which are experimentally difficult to obtain.

In this paper, we study the rupture event of the base sequence
studied in the recent experiments \cite{hatch}.  In Sec. II, we discuss results from 
 atomistic simulations with explicit 
solvent. We shall confine ourselves to a chain of 
12 base-pairs only and focus on the following issues: (i) distribution of stretching of 
hydrogen bonds, and (ii) the extension in the covalent bonds along the strand.
For the comparison, we also study a homosequence (A-T) of same length as studied
in Ref. \cite{Mishra}.Though the distribution obtained here is qualitatively similar to the one obtained in 
Ref. \cite{Mishra}, but showed the asymmetry in the distribution for the both cases.
In Sec. III, we discuss the possible reason for this discrepancy.
Since, the computational cost involved in the atomistic computation is very large and  
beyond our computational limit for a longer DNA, we consider a coarse-grained description of 
dsDNA to study the rupture events.
We incorporate consequences arising due to the semi-microscopic details of nucleotides
in the model to explain the asymmetry in the distributions. In order to substantiate our findings, 
we revisit the ladder model of DNA \cite{degennes} in Sec. IV, and redefine the de Gennes 
characteristic length for a realistic chain to obtain the modified formula for the rupture force, 
which is in good agreement with the coarse-grained simulations.  Analytical results 
developed here are consistent with the experiments and simulations. 
Finally, Sec. V concludes with a brief discussion.

\section{Atomistic Simulations of DNA Rupture}
\begin{figure}[t]
\includegraphics[width=4.5in]{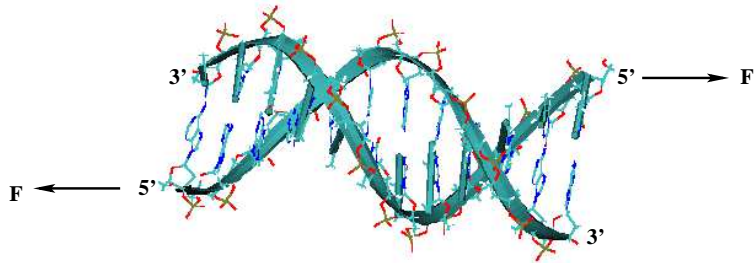}
\caption{Schematic representation of dsDNA under the shear force applied at $5' - 5'$ ends.}
\label{fig-3}
\vspace {0.5cm}
\end{figure}

In order to have a better understanding of the rupture events, we 
perform  atomistic simulations of experimental \cite{hatch} 
sequence of length of 12 base-pairs with explicit solvent.
More specifically, here, we are interested in the distribution of extension in 
the hydrogen and covalent bonds of dsDNA along the chain at the semi-microscopic level,
which is otherwise difficult to obtain.
\begin{figure}[t]
\includegraphics[width=6in]{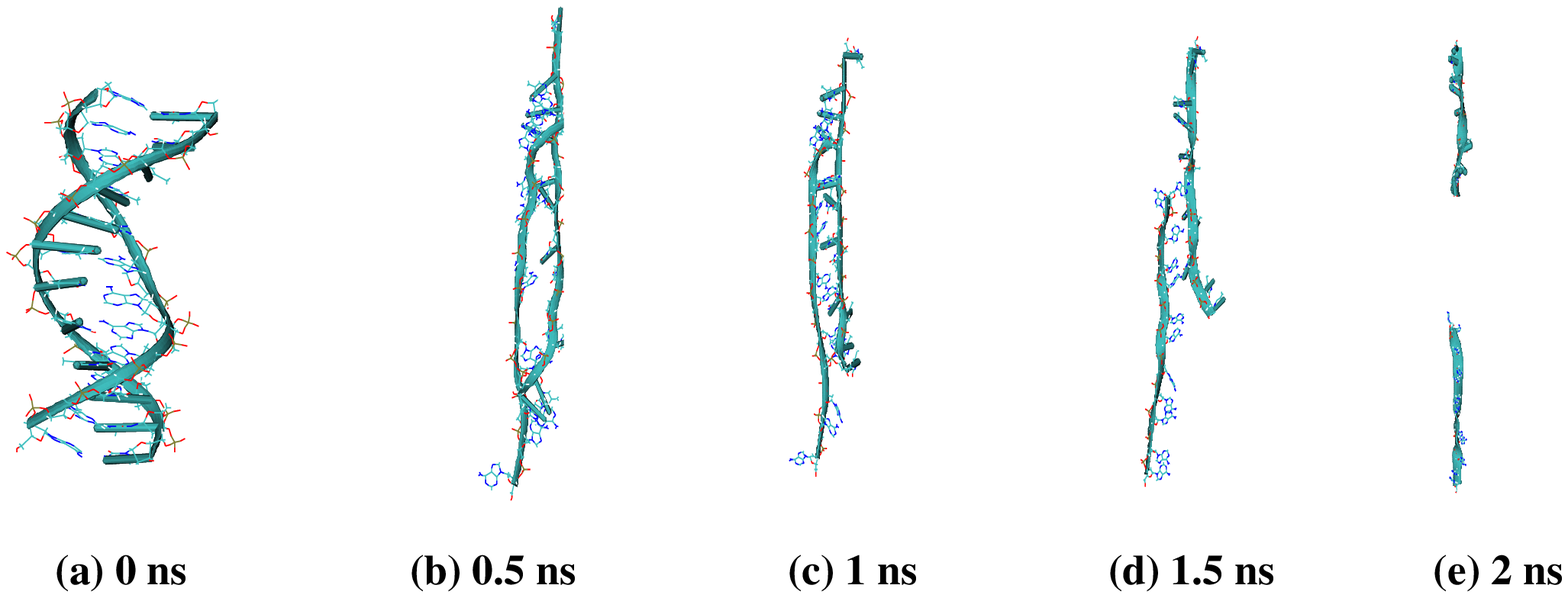}
\label{fig-4}
\vspace {0.5cm}
\caption{Snapshots (generated using VMD software) of dsDNA ($N = 12$)  under constant shear force 
applied at $5'- 5'$ ends, taken at  different time: (a) 0 $ns$, (b) 0.5 $ns$, (c) 1.0 $ns$, 
(d) 1.5 $ns$, and (e) 2.0 $ns$.  These snapshots show the rupture process  at
$T = 320$K.  We have not shown water molecules and counterions in the snapshots for the clarity. }
\end{figure}
We have used  AMBER10 software package \cite{case} with all atom (ff99SB) 
force field \cite{daun} to simulate the rupture event of DNA.  A force routine 
has been added in  AMBER10 to do simulation at constant force \cite{santosh,text}.  
In this case, the force has been applied at
$5'-5'$ ends as shown in Fig.1.  The electrostatic interactions have been  calculated
with Particle Mesh Ewald (PME) method \cite{darden,essmann} using a cubic B-spline interpolation of
order 4 and a $10^{-5}$ tolerance is set for the direct space sum cut off. A real space cut off of
10 $\AA$ is used for both the van der Waal  and the electrostatics interactions.
The starting structure of the DNA duplex sequence (GTCACCTTAGAC) is built using the NAB module
of the AMBER10 suit of  programs. Using the LEaP module in AMBER, we add the $Na^{+}$ (counterions) to
neutralize the negative charges on phosphate backbone group of DNA structure. This
neutralized DNA structure is immersed in water box using TIP3P model for water \cite{Jorgensen}. 
We have 
chosen the box dimension in such a way that the ruptured DNA structure remains fully inside the water
box. For the 12 base-pairs sequence, we have taken the box size of $55 \times 56 \times 199$$\AA^3$
which contains 16690 water molecules and 22  $Na^{+}$ (counterions). The system is 
equilibrated at $F =0 $ for 100 ps under a protocol described in Ref. \cite{maiti,nanolett}. 
We carried out  simulations in the isothermal-isobaric (NPT) ensemble using a time step of 
1 fs. We maintain the constant pressure by isotropic position scaling \cite{case} with a 
reference pressure of 1 atm and a relaxation time of 2 ps.  Constant temperature was maintained 
at 320 K using Langevin thermostat with a collision frequency of 1 ps$^{-1}$. We have used 
$3D$  periodic boundary conditions  during the simulation. 

\begin{figure}[t]
\includegraphics[width=4.8in]{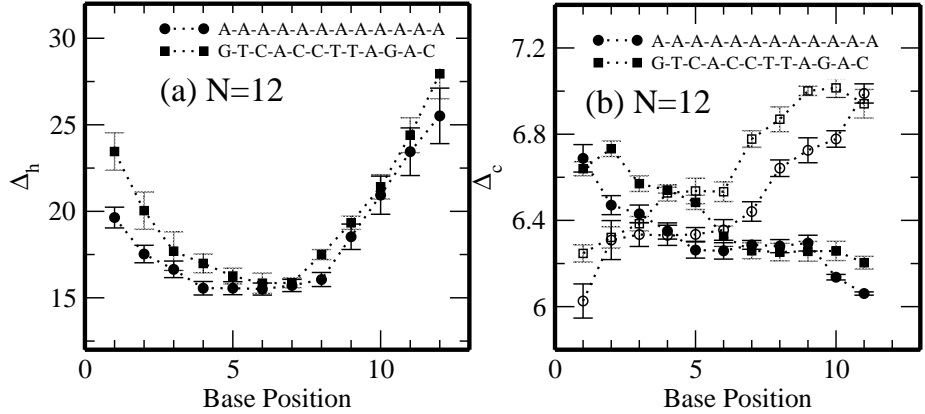}
\caption{Fig (a) shows the variation of extension in hydrogen bond length ($\Delta_h$)
along the chain  with base position for the chain of 12 base-pairs of the 
designed and experimental sequence; (b) Same as Fig. a but for the extension in 
covalent bonds ($\Delta_c$) along the chain length.
Here, open and filled
 symbols correspond to one strand and its complementary strand, respectively. The
 dotted lines are guide to the eye.
}
\label{fig-3}
\vspace {0.5cm}
\end{figure}

To simulate the stretching of hydrogen bonds, we give sufficient time for equilibrium at 
constant force. The magnitude of the applied constant force is 570 pN for the 12 base-pairs,
which is sufficient enough for separating the both strands of dsDNA. To have a better 
understanding, we have monitored the deformations in DNA at different instants of time. 
In Fig. 2, we have shown some of the snapshots of the conformation (generated by visual
molecular dynamics (VMD) software \cite{vmd} under constant 
shear force applied at $5'- 5'$ ends at a temperature 320 K. Initially, the dsDNA remains in 
the zipped state as shown in Fig. 2 (a).   
As time passes, the dsDNA goes to a ladder form (Fig. 2(b)) and then complete rupture takes 
place (Fig. 2 (e)). 
Fig. 3a shows the variation of extension in hydrogen bond length ($\Delta_h$) and 
covalent bond length  ($\Delta_c$) (i.e. deviation from their mean length) along the chain  with base position. 
We have monitored the distance of $C4^{\prime}$ atom of complementary bases in dsDNA to measure 
the extension in hydrogen bond length.  We have studied the system just before the rupture.
Simulation results  show the asymmetry in the distribution of stretching of hydrogen bond
(Fig. 3a) and covalent bonds (Fig. 3b).

In order to see, whether this asymmetry is because of heterogeneity of the sequence,
we have repeated the simulation for a designed homosequence (AAAAAAAAAAAA) of the same length.
The rupture force for this sequence is about 460 pN.
It may be noted that the simulation carried out by Mishra et al \cite{Mishra} or the analytical
solution proposed by de Gennes \cite{degennes}, showed that distribution is symmetric
for a homosequence chain. Surprisingly, even for a homosequence, one can notice that  the extension in 
hydrogen bonds is  more stretched at the pulling end consists of thymine  than the  end 
consists of adenine in the atomistic simulations. The distribution of extension in covalent bonds along 
the chain also shows the asymmetry. One can observe  that the bonds near the pulling end 
( $5^{\prime}$-end) are more stretched and gradually decreases as one approaches the other 
end (i.e., the $3'-$ end).

\section{Coarse-Grained description of DNA Rupture}
One of the possible reasons for such a discrepancy in theoretical models
\cite{degennes,nelson,Mishra} is that they incorporate the same elasticity for the 
both strands. In recent years, there are considerable studies \cite{libchaber,
seol,ke,mishra_pre,kumar_sm,kulveer} on the nature of the elasticity of ssDNA strands.
For example, the force-extension curves of ssDNA (or RNA) consisting of similar type of nucleotides
show the striking differences \cite{seol,ke}. It was found that the poly(T) (or poly U)
show the entropic response whereas  poly(A) show the plateaus arising due to the base 
stacking. These studies provide unequivocal support for the use of different elastic 
constants for complementary strands (say adenine and thymine) in the model.

\begin{figure}[t]
\includegraphics[width=4.0in]{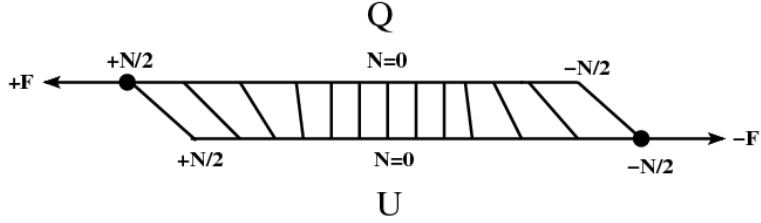}
\caption{Schematic representation of dsDNA under a shear force 
applied at the opposite ends ($5'-5'$ or $3'-3'$). 
}
\label{fig-1}
\end{figure}

In view of this, we now use the coarse-grained description \cite{kumarphys,Li,Kouza,MSLi_BJ07,janke} of 
the flexible polymer chain to model a dsDNA, which allows us to study a system of comparatively
larger  size.  A chain in the model consists of bead units connected by 
effective bonds characterized by the stiff springs (Fig. 4). Each effective bond 
consists of  several chemical bonds ({\it e.g.} sugar phosphate etc. ).
A Lennard-Jones (LJ) potential is used to model the base
pairing interaction  between complimentary nucleotides. The energy of the model 
system is given  by \cite{Li,Kouza,MSLi_BJ07}
\begin{eqnarray}
& & E   =  {\sum_{l=1}^2\sum_{j=1}^N}k^{(l)}({\bf r}_{j+1,j}^{(l)}-d_0)^2
+{\sum_{l=1}^2\sum_{i=1}^{N-2}\sum_{j>i+1}^N}4\left(\frac{C}{{{\bf r}_{i,j}^{(l)}}^{12}}\right) \nonumber \\
 & & + {\sum_{i=1}^N\sum_{j=1}^N}4\left(\frac{C}{(|\bf r_i^{(1)}-\bf r_j^{(2)})|^{12}}-
\frac{A}{(|\bf r_i^{(1)}-\bf r_j^{(2)}|)^6}\delta_{ij}\right),
\end{eqnarray}
where $N$ is the number of beads in each strand. $\bf r_i^{(l)}$ represents the position of $i^{th}$ bead
on $l^{th}$ strand. In present case, $l=1(2)$ corresponds to first (complimentary)
strand of  dsDNA.
The distance between intra-strand beads, $\bf r_{i,j}^{(l)}$, is defined as
$|\bf  r_i^{(l)}-\bf r_j^{(l)}|$. The simplest approach to include semi-microscopic  effects of
nucleotides is to include different elastic constants as discussed above.
The harmonic (first) term with spring constant $k^{l}$ ($k^{1} = Q = 100$  
$\&$   $k^{(2)} = U = 60$) couples the adjacent beads along the two strands.
Second term takes care of excluded volume effect {\it i.e.} two beads
can not occupy the same space \cite{book}.
The third term, described by Lennard-Jones (LJ) potential, takes care of the
mutual interaction between two strands.
The  first term of LJ potential (same as second term of Eq.2) will
not allow the overlap of two strands. Here, we set $C = 1$ and $A=1$.
The second term of LJ potential corresponds to the base-pairing between
two strands. The base-pairing interaction is restricted to the native contacts
($\delta_{ij}=1$) only {\it i.e.}  $i^{th}$ base of $1^{st}$ strand forms pair with
the $i^{th}$ base of $2^{nd}$ strand only as shown in Fig. 4, 
which is similar  to the G\=o model \cite{go}.
The parameter $d_0 (=1.12)$ corresponds to
the equilibrium distance in the harmonic potential, which is close to the
equilibrium position of the LJ potential. In Eq. 2,
we use dimensionless distances and energy parameters \cite{real}. 
The major advantage of this model is that the ground 
state conformation is known. Therefore, equilibration is not an issue here, if
one wants to study the dynamics under the applied force at low $T$ \cite{Li}.
The equation of motion is obtained from the following Langevin equation 
\cite{Allen,Smith,Li,MSLi_BJ07} 

\begin{figure}[t]
\includegraphics[width=2.6in]{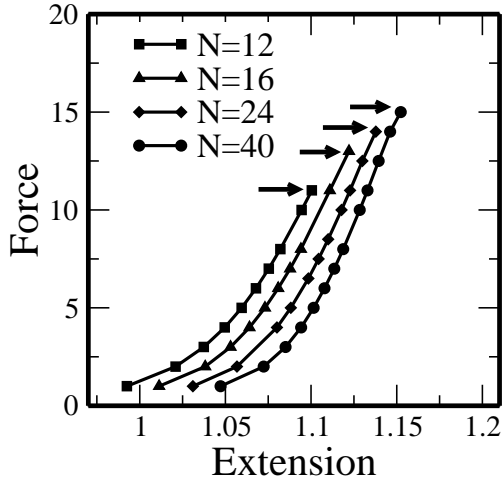}
\caption{Force {\it vs} extension curves for different lengths in the constant 
force ensemble. Arrows indicate the maximum force, where native contacts 
approaches to zero and two chains are separated. For the sake of comparison, we
have normalized the extension by its chain length.
}
\label{fig-6}
\vspace {0.5cm}
\end{figure}
\begin{equation}
m\frac{d^2{\bf r}}{dt^2} = -{\zeta}\frac{d{\bf r}}{dt}+{\bf F_c(t)}+\Gamma(t),
\end{equation}
where $m ( =1 )$ and $\zeta (=0.4)$ are the mass of a bead and the friction 
coefficient, respectively. Here, $F_c$ is defined as $-\frac{dE}{d{\bf r}}$ and 
the random force $\Gamma$ is a white noise \cite{Smith},
i.e., $<{\Gamma(t)\Gamma(t')}>=2\zeta T\delta(t-t')$, which ensures that 
the temperature of the system remains constant during the simulation 
for a given $f$. The $6^{th}$ order predictor-corrector algorithm with time step 
$\delta t$=0.025 \cite{Smith} has been used to integrate the equation of motion.  
These results are averaged over many trajectories. The equilibration has been checked by 
monitoring the stability of data against at least ten times longer run. We have used 
$2\times10^9$ time steps out of which first $ 5\times 10^8$ steps are not 
taken in the averaging.

In the constant force ensemble, we add an energy $-\bf {f}.\bf {x}$ to the 
total energy of the system given by Eq. 2, where $\bf x$ is the extension along the
applied force direction. The force-extension ($f-x$) curve 
is shown in Fig. 5 for different lengths. The $f-x$ curve shows the entropic 
response at low forces and remains qualitatively similar to the one seen in 
experiments. It may be noted that in the ladder model such response is missing
as system remains in the stretched state. The rupture force is defined as a maximum force, 
where  all the native contacts ({\it i.e.} number of intact base-pairs) 
suddenly goes to zero.  The variation of the rupture force as a function of length of the 
chain for the low temperature ($T = 0.06$) is shown in Fig. 6. One can notice
that the rupture force approaches to an asymptotic value as length of the chain 
increases and is consistent with the experiment \cite{hatch}.

\begin{figure}[t]
\includegraphics[width=2.8in]{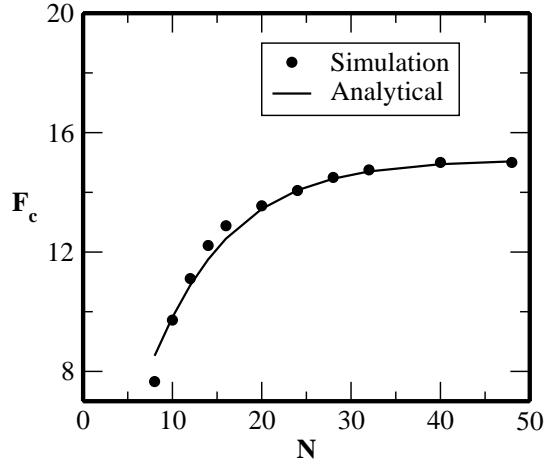}
\caption{The variation of rupture force with chain length. The solid
line corresponds to a fit of Eq.5 with $f_1 =1$ and $\chi \approx 0.118$.
Solid circles represent the value obtained through the simulation. A nice
agreement with the theoretical prediction (Eq. 5) is apparent from the plot.
}
\label{fig-7}
\vspace {0.5cm}
\end{figure}

\begin{figure}[t]
\includegraphics[width=4.5in]{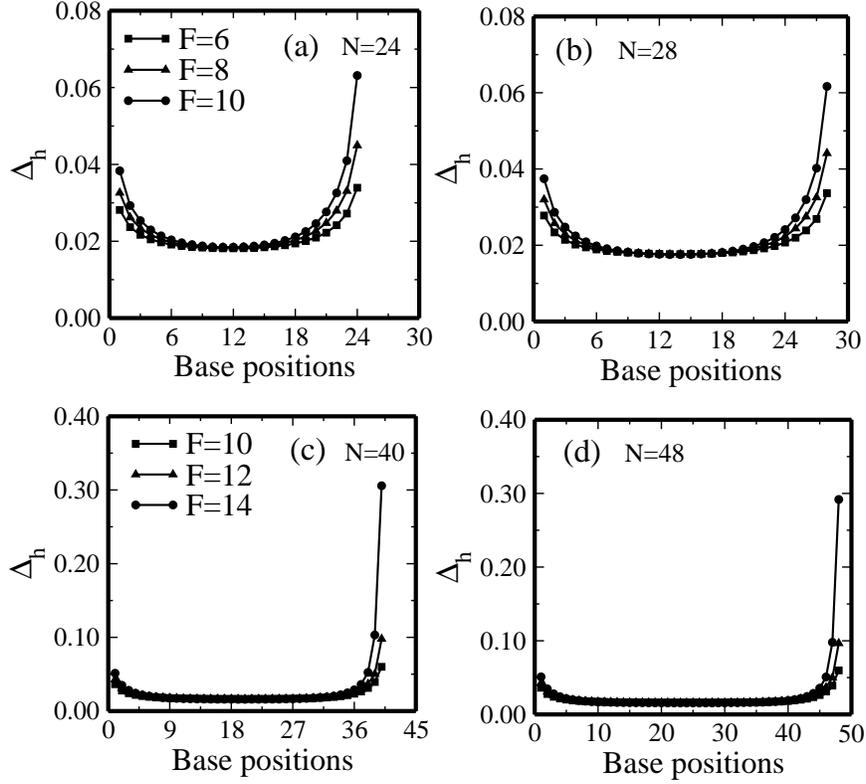}
\caption{ Figs a,b,c \& d show the variation of extension in hydrogen bond 
length ($\Delta_h$) along the chain for the different length 
N=24, 28, 40 and  48 respectively. 
It is obvious from these plots that the extension in hydrogen bonds for each length is
asymmetric. It is because, both strands have different elastic constants
($k^{1}\ne k^{2}$). One can see that the elongation at the end of a strand
having low elastic constant (where the force is applied) is much more than the
middle one, where the differential force approaches to zero.
}

\label{fig-8}
\end{figure}

\begin{figure}[t]
\includegraphics[width=4.5in]{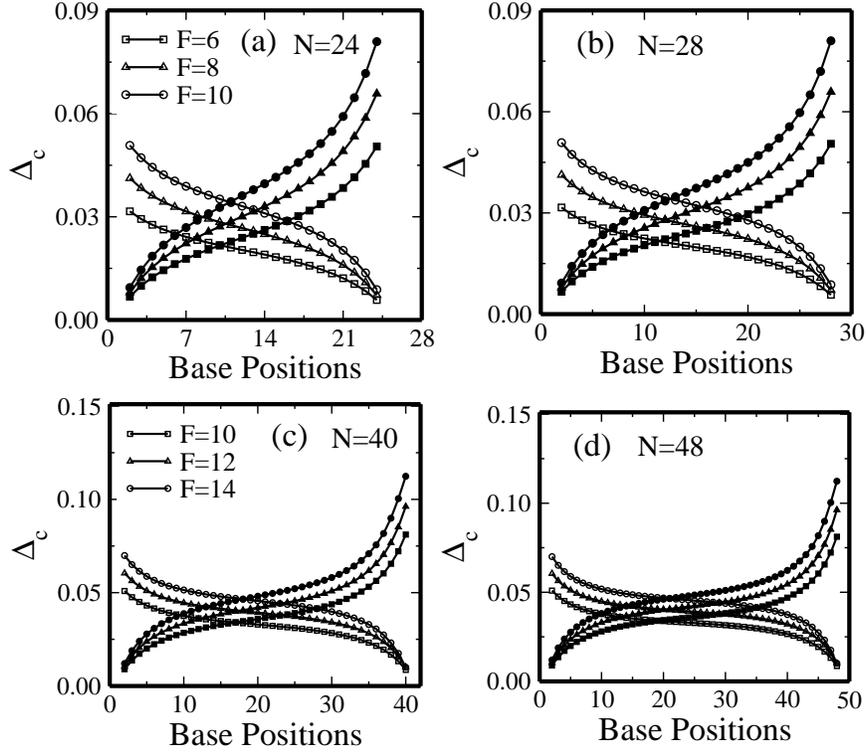}
\caption{ Figs a,b,c \& d show the variation in extension of covalent 
bond length ($\Delta_c$) along the chain for the different chain 
length N=24, 28, 40 and 48, respectively.  
The asymmetry  in the extension of covalent bonds is apparent from these plots 
for each length.
Open and filled symbols correspond to one 
strand and its complementary strand, respectively. The larger extension 
in bond length corresponds to the end of a strand having low elastic constant
(where the  force is applied), whereas the minimum extension in bond length
corresponds that the force is applied at the other end of the complimentary
strand having larger elastic constant. The differential force here also 
approaches to zero.
} 

\label{fig-9}
\end{figure}

In Figs. 7 a, b, c \&  d, we show the distributions of extension in hydrogen bonds ($\Delta_h$)
along the chain for four different lengths. It may be noted that for the 
$k^{1} = k ^{2}$ (or $ Q = U$), the distributions are symmetric \cite{Mishra}. However, for $k^{1} \ne k ^{2}$
(or $Q \ne U$), the distribution is asymmetric, which is consistent with the atomistic simulations 
presented in Sec. II.  
The characteristic de Gennes length for the present simulation appears to be approximately   
8 bases, whose precise value is unknown. 
From this plot, one can observe that the hydrogen bonds near the extreme ends (up to 
$\approx 8$ bases) get stretched, while the bases in the middle above the de Gennes length 
($\approx$ 8 to 30 bases) remain unstretched indicating that the differential shearing force approaches 
to zero in this region. In Figs. 8 a, b, c \&  d, we show the variation of extension in 
the covalent bonds (back bone) along the chain. We observe  similar asymmetry in the extension of 
covalent bonds for all lengths. The curve has three distinctively different regions. 
It shows that bonds near the pulling end (say 5'-end)  are stretched more and 
gradually decrease. 
After the de Gennes length, they  saturate and remain almost the same. However, 
when one approaches the other end (i.e. 3'-end), there is a change in the 
slope and the extension is quite less compare to the middle one. This is because of
the fact that   the  3'-end of first strand is near to the 5'-end of the other chain, 
where a similar force is also applied in the opposite direction. Since, the dsDNA is in 
the zipped state, the applied force at 5'-end of one strand also pulls the other 
strand along the opposite direction, which causes a relatively slower increase. 
Needless to mention that this  increase also approaches to a constant value indicating that the 
differential shearing force also vanishes after the de Gennes length.

\section{ DNA Rupture: Analytical solution}
In order to get the precise value of de Geness length and critical rupture force,
we consider the ladder model of DNA \cite{degennes}. The semi-microscopic details 
{\it e.g.} inter and intra-strand stacking interactions, which give rise helecoidal 
structure, effect of pulling at $3'-3'$ and $5'-5'$  and heterogeneity in the sequence 
have not been included in the model. However, the covalent bonds of two strands (say made up 
of adenine (A) and thymine (T))  have been modeled by harmonic potentials with different
spring constants ($Q \ne  U$). The base-pairing interaction for homosequence 
DNA is also modeled by the harmonic potential with the spring constant $R$. 
We apply the shear force on two strands as shown in Fig. 4. Let the displacements
of the upper strand (say made up of A with large spring constant $Q$) be ${u_n}$ and lower strand 
(made up of thymine with smaller spring constant $U$)  be ${v_n}$ for the $n^{\it th}$ base-pair 
in a DNA chain of length $N$ base-pairs. The Hamiltonian for the chain can be expressed 
as \cite{degennes}

\begin{eqnarray}
H  & = & \sum_{n =-\frac{N}{2}}^{\infty}\frac{1}{2} Q(u_n-u_{n+1})^2 + \sum_{n =-\infty}^{\frac{N}{2}}\frac{1}{2} 
U(v_n-v_{n+1})^2 \nonumber \\ & + &  \sum_{n =-\frac{ N}{2}}^{\frac{N}{2}}\frac{1}{2} R(v_n-u_n) ^2. \label{ham}
\end{eqnarray}
Following Ref. \cite{degennes}, we evaluate the expression for de Gennes characteristic length
 ${\chi^2 = \frac{R(Q+U)}{QU}}$, when the elasticity of two strands are not same. Here, $R << Q$ and $U$.
The relation between  the rupture force $F_C$ and the chain length for this model  is given by 
(see Appendix A) 
\begin{equation}
\frac{F_c}{f_1} = \frac{2\tanh(\chi N/2)}{\chi(1+ \frac{(Q-U)}{(Q+U)}\left[\frac{\chi+\tanh(\chi N/2)}{1+
\chi\tanh(\chi N/2)}\right]\tanh(\chi N/2))}.
\end{equation}
Though, the equation looks complicated, but it has only two free parameters $Q$ and $U$, 
whose value can be obtained experimentally. For a long chain, Eq. 5 reduces to
\begin{equation}
\frac{F_c}{f_1} \approx  \frac{Q+U}{\chi Q}.
\end{equation}
For ${Q=U}$,  Eq. 5 reduces to the Eq. 1 as proposed by de Gennes \cite{degennes}. In Fig. 9, we depict the
distribution of extension in hydrogen bonds and covalent bonds with base position at 
critical force $F_c$ (see Appendix A). These distribution are qualitatively similar to the one obtained in the
atomistic simulations (Sec. II) and the coarse-grained simulations (Sec. III).

\begin{figure}[t]
\includegraphics[width=4.5in]{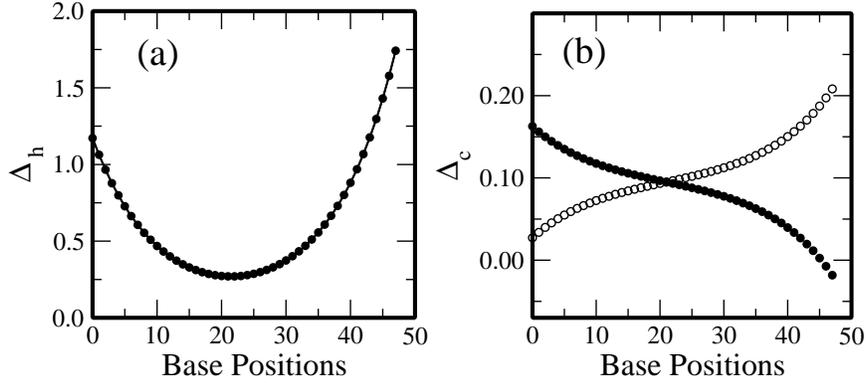}
\caption{Fig (a) shows the variation of extension in hydrogen bond length ($\Delta_h$)
along the chain  with base position. 
(b) Variation in extension of covalent bond length ($\Delta_c$) along the 
chain length. Here, open and
filled symbols correspond to one strand and its complementary strand, respectively.
}

\label{fig-2}
\vspace {0.5cm}
\end{figure}

The de Gennes characteristic length for the L-J potential can be obtained
by expanding  the L-J potential given in Eq.2 around its equilibrium value and 
equating the coefficient of second term of its expansion with harmonic spring.
For the present model it is estimated to be $\approx 0.118$. Substituting values of 
$f_1 (=1)$ and the above mentioned value of $\chi$ in Eq. 5, one can obtain the value of 
$F_c$ for a given length of dsDNA.  In Fig. 6, we also show the behavior of the 
rupture force as a function of DNA length obtained from Eq. 5. One can notice a
nice agreement with simulations and analytical result obtained here.

\section{Conclusions}

In this paper, we have performed atomistic simulations to study the effect of shear force 
on the rupture of dsDNA. In contrast to the previous studies \cite{degennes,nelson,Mishra},
here the distribution of extension in hydrogen and covalent
bonds show the asymmetry. This asymmetry arises because of different elastic constants
of the strands. Inclusion of different elastic constants in the description of the 
coarse-grained model gives qualitatively similar behavior as seen in the atomistic simulations. 
For a short chain, we find that the rupture force increases linearly and saturates for a longer
chain, which is consistent with the experiment and earlier studies \cite{hatch, Mishra}. 
The distribution of hydrogen bonds show that differential force penetrates up to the 
de Gennes characteristic length. Using the ladder model of DNA, we have obtained the analytical
expression for the de Gennes characteristic length and rupture force for the chains
whose complimentary strands have different elastic constants. These values are in very
good agreement with the coarse-grained simulations. By setting $k^1 = k^2$ (or $Q = U$), 
the expression reduces to the de Gennes expression Eq. 1 \cite{degennes}.

It is possible to extend the approach developed here in understanding many intramolecular processes 
such as microsatellites formation, bulge loop propagation in repetitive sequences
which exhibits complex dynamics and a distinct biological function. Our studies may
provide the mechanism involved in ligand receptor binding in cell's tissue
at molecular level and DNA protein interactions. An all atoms simulation can provide the life time of 
these interactions, which can be verified by the time resolved spectroscopy. One would also 
able to know whether all interactions contribute at the same moment or  have different 
life times.

We thank Garima Mishra, M. Santosh and P. K. Maiti for many helpful discussions on the subject. 
We also thank P. K. Maiti for providing us the constant force routine for the atomistic simulations.
Financial supports from the Department of Science and Technology, Council of 
Scientific and Industrial Research, New Delhi are gratefully acknowledged. The
generous computer support from IUAC New Delhi is also acknowledged by the
authors. One of the authors (SK) would like to acknowledge the Sonderforschungsbereich/Transregio 
SFB/TRR 102 Halle-Leipzig {\em Polymers under Multiple Constraints: Restricted and Controlled 
Molecular Order and Mobility\/} for his visit at Institute of Physics, University of Leipzig, Germany,
where a part of the work is carried out.

\appendix
\section{}
Under the application of shear force,  Eq. 4 gives the equilibrium condition for the upper strand as
\begin{equation}
\frac{\partial H}{\partial u_n} = Q(u_{n+1}-2u_n+u_{n-1}) + R(v_n-u_n) = 0
\end{equation}
and similarly for the lower strand,
\begin{equation}
\frac{\partial H}{\partial v_n} = U(v_{n+1}-2v_n+v_{n-1}) + R(u_n-v_n) = 0.
\end{equation}
For large N, Eq.A.1 and Eq.A.2 can be expressed in continuum limit of $n$ as 
\begin{eqnarray}
Q\frac{d^2u_n}{dn^2} + R(v_n-u_n) &=& 0\\
U\frac{d^2v_n}{dn^2} - R(v_n-u_n) &=& 0.
\end{eqnarray}
By adding Eqs. A.3 and A.4, we get
\begin{equation}
Q\frac{d^2u_n}{dn^2} + U\frac{d^2v_n}{dn^2}=0.
\end{equation}
Thus from the solution of Eq. A.5  as the total tension constant, we obtain the following condition:
\begin{equation}
Qu_n + Uv_n = nF.
\label{ten}
\end{equation}
On multiplying Eq. A.3 by U and Eq. A.4 by Q, and subtracting we obtain,
\begin{eqnarray}
 \frac{d^2\delta_n}{dn^2} - \frac{R(Q+U)}{QU}\delta_n = 0,
\end{eqnarray}
where, $\delta_n = v_n - u_n$.
This is a simple second order differential equation whose solution is of the form:
\begin{equation}
\delta_n = \delta_0\cosh(\chi n)  + A\sinh(\chi n), 
\end{equation}
where ${\chi^2 = \frac{R(Q+U)}{QU}}$ and A is an arbitrary constant of integration.
From Eqs. A.6, A.7 and A.8, we get
\begin{eqnarray}
v_n & = &  \frac{nF}{Q+U} + \frac{Q}{Q+U}\delta_0\cosh(\chi n) \nonumber   \\ 
& + &  \frac{AQ}{Q+U}\sinh(\chi n)
\end{eqnarray}
\begin{eqnarray}
u_n & = & \frac{nF}{Q+U} - \frac{U}{Q+U}\delta_0\cosh(\chi n)   \nonumber \\
&  - &    \frac{AU}{Q+U}\sinh(\chi n)
\end{eqnarray}
The force at both the ends ($\frac{N}{2}$ and $-\frac{N}{2}$) of the strand must be balanced.
Thus,
\begin{eqnarray}
F &=& U(v_{\frac{N}{2}} - v_{\frac{N}{2}-1}) + R(v_{\frac{N}{2}}-u_{\frac{N}{2}})\\
-F &=& Q(u_{-\frac{N}{2}}-u_{-(\frac{N}{2}-1)})+R(u_{-\frac{N}{2}}-v_{-\frac{N}{2}}),
\end{eqnarray}
which gives a relation between ${A}$ and ${\delta_0}$ as
\begin{equation}
A = \delta_0\frac{(Q-U)}{(Q+U)}\frac{(\sinh(\chi N/2) + \chi \cosh(\chi N/2))}{(\cosh(\chi N/2) + \chi \sinh(\chi N/2))}
\end{equation}
The overall force acting on the base-pairs of the dsDNA can be calculated
as the sum of restoring forces on the base-pairs,
\begin{eqnarray}
F &=& \sum_{n=-\frac{N}{2}}^{\frac{N}{2}}R\delta_n = \int_{-N/2}^{N/2}R \delta_n dn = \frac{2R\delta_0 \sinh(\chi N/2)}{\chi}
\end{eqnarray}
The rupture will take  at critical force $F_c$ from the end $({n=N/2})$, because the end base-pair ($N/2$)will have maximum 
elongation. If the force required to break a base-pair is ${f_1}$, then
\begin{eqnarray}
f_1 &  = & R\delta_0 \left [\cosh(\chi N/2) + \right .\nonumber \\
 & +  & \left .\frac{(Q-U)}{(Q+U)}\left[\frac{\sinh(\chi N/2)  + \chi \cosh(\chi N/2)}{\cosh(\chi N/2) + 
\chi\sinh(\chi N/2)}\right] \sinh(\chi N/2)\right ]  \nonumber \\
\end{eqnarray}
Dividing Eq. A. 14 by A.15, we get a relation between the rupture force $F_C$ and the ${f_1}$
\begin{equation}
\frac{F_c}{f_1} = \frac{2\tanh(\chi N/2)}{\chi(1+\frac{(Q-U)}{(Q+U)}\left[\frac{\chi+\tanh(\chi N/2)}{1+
\chi\tanh(\chi N/2)}\right]\tanh(\chi N/2))}
\end{equation}


\begin{thebibliography}{99}
\bibitem{real} For melting the free energy per base is approximately -1.2 kcal/mol, which includes
base pairing as well as stacking (inter and intra). However in rupture, we assume that there 
are breaking of hydrogen bonds (base pairing) only and stacking does not contribute significantly. 
For such a case if one takes free energy approximately -0.6 kcal/mol per base of the zipped 
conformation \cite{Mishra} and equate it with the ruptured state, it is possible to parametrize
the rupture force also.
\end{thebibliography}
\end{document}